\documentclass[twocolumn,secnumarabic,amssymb, nobibnotes, aps, prd]{revtex4-1}
\usepackage{graphicx}
\usepackage[T1]{fontenc}
\usepackage[utf8]{inputenc}
\usepackage[colorlinks,linkcolor=blue,anchorcolor=blue,citecolor=blue]{hyperref}
\usepackage{indentfirst}
\usepackage{amsmath}
\usepackage{subcaption} 
\begin{document}

\title{Calculations of Di-Hadron Production via Two-Photon Processes in Relativistic Heavy-Ion Collisions}
\author{Luobing Wang}
\affiliation{State Key Laboratory of Particle Detection and Electronics, University of Science and Technology of China, Hefei 230026, China}
\author{Xinbai Li}
\affiliation{State Key Laboratory of Particle Detection and Electronics, University of Science and Technology of China, Hefei 230026, China}
\author{Zebo Tang}
\affiliation{State Key Laboratory of Particle Detection and Electronics, University of Science and Technology of China, Hefei 230026, China}
\author{Xin Wu}
\affiliation{State Key Laboratory of Particle Detection and Electronics, University of Science and Technology of China, Hefei 230026, China}
\author{Wangmei Zha}
\email[Corresponding author, ]{Wangmei Zha Address: No. 96 Jinzhai Road, Hefei city, China Tel: +86 551 63607940  Email: first@ustc.edu.cn}
\affiliation{State Key Laboratory of Particle Detection and Electronics, University of Science and Technology of China, Hefei 230026, China}

\begin{abstract}

Two-photon processes in relativistic heavy-ion collisions have emerged as a critical probe of quantum electrodynamics in ultra-intense electromagnetic fields, with recent focus extending beyond dileptons to hadronic final states. At present, quantitative studies of di-hadron production via two-photon interactions remain scarce. In this work, we employ the Equivalent Photon Approximation and the two-photon fusion measurements from \(e^{+}e^{-}\) collisions to obtain differential cross-section predictions for \(\pi^{+}\pi^{-}\), \(K^{+}K^{-}\), and \(p\bar{p}\) pairs produced in ultra-peripheral \(\mathrm{Au{+}Au}\) collisions at \(\sqrt{s_{NN}} = 200\,\text{GeV}\) within the STAR acceptance, as well as in \(\mathrm{Pb{+}Pb}\) collisions at \(\sqrt{s_{NN}} = 5.36\,\text{TeV}\) within typical LHC acceptance. The calculations deliver the unified baseline for light-meson and baryon pairs in this environment, supplying benchmarks for upcoming STAR and LHC measurements and guiding future systematic investigations of hadronic two-photon processes at RHIC and LHC facilities.

\end{abstract}

\keywords{Ultra-peripheral collisions \and Two-photon processes \and Equivalent photon approximation \and Di-hadron production}
\maketitle

\section{Introduction} 
\label{sec::Intro}

Relativistic heavy-ion collision experiments, at RHIC and LHC, are primarily motivated by the search and characterization of the quark-gluon plasma (QGP), a state of deconfined quarks and gluons at extreme temperature and density~\cite{QGP_Search}. In such collisions, the large charge $Z$ of the fast-moving nuclei also generates enormous electromagnetic fields that can be treated as a flux of quasi-real photons~\cite{QED_in_UPC,KRAUSS1997503}. These photon-induced interactions are especially prominent in ultra-peripheral collisions (UPC), where the nuclear impact parameter $b$ exceeds the sum of the nuclear radii and hadronic interactions are suppressed. Beyond QGP-related phenomena, UPC events provide a unique opportunity to study two-photon ($\gamma \gamma$) and photon-nuclei ($\gamma A$) processes in a new regime using heavy-ion projectiles.

The concept of treating the Coulomb field of a relativistic charged projectile as a spectrum of virtual photons dates back to Enrico Fermi in 1924~\cite{Fermi_2003}.  A decade later, Weizsäcker and Williams independently developed this idea into the formalism of the Equivalent Photon Approximation (EPA)~\cite{weizsacker1934ausstrahlung,williams1934nature}.  In the EPA, the time-varying electromagnetic field of a relativistic nucleus is factorized into an equivalent flux of photons carrying a distribution of energies.  This approach allows one to calculate the reaction cross sections induced by $\gamma$ by integrating the effective photon flux with the pertinent elementary cross section $\gamma$ -target. The EPA  is most transparent in ultra-peripheral collisions, where hadronic interactions are largely absent. The same formalism can also describe photon-induced production that persists when the nuclei still overlap hadronically \cite{UPCreview,UPCreview2}. Together, these frameworks underpin modern calculations of photon–photon and photon–nucleus reactions in heavy-ion environments.

Experimentally, two-photon interactions in heavy-ion collisions have been studied most extensively via dilepton production. The simplest such process is $\gamma\gamma \to \ell^+\ell^-$ (the process of matter-antimatter creation), which is a pure QED phenomenon. Early measurements at RHIC~\cite{STAR_ChenJH} demonstrated substantial $e^+e^-$ pair production in Au+Au UPC, consistent with QED theoretical expectations for photon fusion (Breit–Wheeler process)~\cite{Star2004:e+e-,Star2017:e+e-,PHENIX：B-W}. At LHC, higher energies have enabled the copious production of electron and muon pairs in Pb+Pb UPC events, observed by ALICE, ATLAS, and CMS~\cite{alice2013dielectron,atlas2023diele,CMS:2021dimuon,CMS_2024_diel_NLO,ATLAS_2020_dimuon}. These exclusive dilepton final states are identified by two opposite-charge leptons with no other particle activity, and their kinematic distributions (e.g., pair acoplanarity and $p_T$) have been shown to agree with QED predictions for $\gamma\gamma$ fusion~\cite{alice2013dielectron,atlas2023diele,CMS:2021dimuon,Star2004:e+e-,Star2017:e+e-}. Recently, even the production of the much heavier $\tau^+\tau^-$ pairs via photon–photon collisions has been observed at the LHC~\cite{cms2023ditao,atlas2023ditao}, further extending this program. Together, these results confirm that the intense electromagnetic fields of relativistic ions can induce photon–photon reactions analogous to those long studied in $e^+e^-$ colliders, providing a novel testbed for quantum electrodynamics in the strong-field regime.

Given this success with dilepton channels, there is growing interest in extending two-photon studies to purely hadronic final states. In particular, the production of meson or baryon pairs via $\gamma\gamma$ fusion in UPC opens a new category of processes that can be investigated. Although the ALICE Collaboration and LHCb Collaboration has reported the observation of $K^+K^-$ pairs in Pb+Pb ultra-peripheral collisions at the LHC~\cite{Alice_KpKm,LHCb_KpKm}, the production is predominantly driven by the $\gamma + A$ process, while the $\gamma\gamma$ contribution is negligible~\cite{Alice_KpKm}. The STAR Collaboration has recently reported the observation of exclusive proton–antiproton pair production in Au+Au ultra-peripheral collisions ($\gamma\gamma \to p\bar{p}$)~\cite{Wu:QM2025}. However, in contrast to the dilepton case, there is a general lack of theoretical calculations and predictions available for such di-hadron production channels in heavy-ion collisions. Only very few studies have attempted to calculate two-photon hadron-pair yields in the UPC environment~\cite{ppbar_Pu,ppbar_Shao,ppabr_klusek,dipion_fit}, and no comprehensive framework exists yet to directly compare with measurements like the new $p\bar{p}$ result. This paucity of predictions underlines the importance of developing models for $\gamma\gamma \to$ hadrons in nuclear collisions, and of benchmarking them against empirical data. In this work, we present an EPA-based calculation of di-hadron production in Au+Au ultra-peripheral collisions at $\sqrt{s_{NN}}=200$ GeV, focusing on the exclusive channels $\pi^+\pi^-$, $K^+K^-$, and $p\bar{p}$. 
In addition, we also evaluate the corresponding production in \(\mathrm{Pb{+}Pb}\) collisions at \(\sqrt{s_{NN}} = 5.36~\mathrm{TeV}\), covering the same exclusive channels.

The input photon–photon cross-sections for these hadronic final states are constrained using existing $e^+e^-$ collision data on $\gamma\gamma \to$ di-hadron~\cite{Belle:2005fji,CLEO:1993ahu,TASSO:1983mvj,JADE:1986ezn,OPAL:2002nhf,VENUS:1997can,L3:2003gyz,ARGUS:1988apz,dikaon_1_LEP,dikaon_2_dipion_1_ALEPH,dikaon_3_argus,dikaon_4_belle2003,dikaon_5_dipion_2_belle2005,dikaon_6_dipion_3_CLEO,dikaon_7_TASSO,dikaon_8_dipion_4,dipion_5_SPEAR,dipion_6,dipion_7_MRAKII,dipion_8,dipion_9_Belle2007,dipion_11,dipion_12,dipion_13}. In these $e^{+}e^{-}$ measurements, the photon virtualities are typically controlled by operating in an untagged configuration, in which neither the scattered electron nor positron is detected outside the beam pipe. In heavy-ion ultra-peripheral collisions, the quasi-real nature of the equivalent photons instead follows from the requirement that the photon emission be coherent over the entire nuclear charge, which constrains the transverse momentum to be of the order of $p_{T}\lesssim \hbar /R$~\cite{baur1990coherent}, where $R$ denotes the nuclear radius. As a consequence, the typical photon virtualities are of the order of $Q^{2}\sim10^{-2}\,\mathrm{GeV}^{2}$ in untagged $e^{+}e^{-}$ measurements, while they are further suppressed to $Q^{2}\lesssim10^{-3}\,\mathrm{GeV}^{2}$ in heavy-ion UPC for gold and lead nuclei~\cite{Wu_Q}. In the prevailing view, the photons in the \(e^{+}e^{-}\) and heavy-ion UPC systems discussed here are considered quasi-real, and their properties are regarded as nearly identical to those of real photons. Consequently, the effects arising from the differences in photon virtualities between the two systems are generally expected to be negligible. However, alternative theoretical considerations suggest that such a change in photon virtuality may lead to modifications of the cross section by up to an order of magnitude, as predicted in certain models~\cite{Qdependence}.

We fold these elementary cross-sections with the appropriate photon flux distributions from each Au nucleus (via the EPA) to obtain production rates in heavy-ion collisions. Furthermore, the calculations incorporate the relevant kinematic acceptance and selection criteria of the STAR and LHC experiments, respectively, enabling direct comparison with their respective measurements. The results provide quantitative predictions for two-photon hadron-pair yields in 200 GeV Au+Au UPC and 5.36 TeV Pb+Pb UPC, which can be confronted with upcoming experimental data and thereby help fill the gap in understanding $\gamma\gamma$ processes beyond the dilepton sector.

\section{METHODOLOGY}

The cross section for the electromagnetic production of a two-photon process in heavy-ion collisions can be calculated based on a classical equivalent photon distribution ~\cite{KRAUSS1997503}. The fusion cross section can be obtained by folding the quasi-real photon fluxes from the two colliding nuclei with the cross section for the process \(\gamma \gamma \rightarrow X\), where experimental data are utilized in the specific calculations. 

\begin{equation}
\begin{split}
\sigma_{A_1 + A_2 \rightarrow A_1 + A_2 + X}
&= \int_{0}^{\infty} d\omega_1 \, n(\omega_1) \int_{0}^{\infty} d\omega_2 \, n(\omega_2)\, \\
&\quad \times\sigma_{\gamma \gamma \rightarrow X}(W) ,
\end{split}
\end{equation}
where \( A_1 \) and \( A_2 \) denote the colliding nuclei, \( \omega_1 \) and \( \omega_2 \) represent the equivalent photon energies, and \( n(\omega_1) \) and \( n(\omega_2) \) correspond to the equivalent photon flux per unit energy at energies \( \omega_1 \) and \( \omega_2 \), respectively. The term \(\sigma_{\gamma \gamma \rightarrow X}\) represents the cross section for the reaction \(\gamma \gamma \rightarrow X\), which is dependent on the invariant mass \( W \) of the produced final state \( X \).

However, the classical equivalent photon distribution integrates the photon flux over the entire transverse plane, thereby concealing the information about the photon flux at specific distances from the nucleus center. To improve accuracy, the dependence of the photon flux on the impact parameter, as well as the shielding effects due to hadronic interactions, must be considered. When projected onto the beam direction, the vectors from the centers of nuclei \( A_1 \) and \( A_2 \) to a point in transverse space are defined as \(\mathbf{b}_1\) and \(\mathbf{b}_2\), respectively. A straightforward geometrical relationship is established:

\begin{equation}
    \begin{split}
    \mathbf{b}_{1} - \mathbf{b}_{2} = \mathbf{b}.
    \end{split}
\end{equation}
With this in mind, we can derive the precise form of the production cross section for the two-photon process in ultra-peripheral collisions (UPC):
\begin{equation}
\begin{split}
\sigma_{A_{1} + A_{2} \rightarrow A_{1} + A_{2} + X} 
&= \int d^{2}\mathbf{b}\int d^{2}\mathbf{b}_{1}\int d\omega_1\int d\omega_2 \\
&\quad \times N(\omega_{1}, \mathbf{b}_{1})\, N(\omega_{2}, \mathbf{b}_{2}) \\
&\quad \times P_{\mathrm{NH}}(b)\, \sigma_{\gamma \gamma \rightarrow X}(W),
\end{split}
\end{equation}
where \(N(\omega_{1}, \mathbf{b}_{1})\)is the photon flux per unit energy and per unit area emitted by nucleus  \(A_{1}\) at transverse distance with photon energy \(\omega_{1}\) in the beam view, and \(N(\omega_{2}, \mathbf{b}_{2})\) is the corresponding photon flux for nucleus \(A_{2}\). The integrations over $\mathbf{b}$ and $\mathbf{b}_{1}$ extend over the entire two-dimensional transverse plane in beam view. The integration over the impact parameter $\mathbf{b}$ accounts for all possible collision geometries, while for a fixed $\mathbf{b}$, the convolution over the transverse coordinates incorporates all possible two-photon production configurations.

Upon the change of variables from $(\mathbf{b}, \mathbf{b}_1)$ to $(\mathbf{b}_1, \mathbf{b}_2)$, and introducing the angle $\phi$ between $\mathbf{b}_1$ and $\mathbf{b}_2$, the integral can be simplified as follows\cite{cahn1990}:
\begin{equation}
     \begin{aligned}
        \frac{d\sigma_{A_{1} + A_{2} \rightarrow A_{1} + A_{2} + X}}{d\omega_{1} d\omega_{2}} &= \int_{0}^{\infty} 2\pi b_{1} db_{1} \int_{0}^{\infty} 2\pi b_{2} db_{2} \\
        &\quad \times \int_{0}^{2\pi} \frac{d\phi}{2\pi} N(\omega_{1}, \mathbf{b}_{1}) N(\omega_{2}, \mathbf{b}_{2}) \\
        &\quad \times P_{\mathrm{NH}}(b) \sigma_{\gamma \gamma \rightarrow X}(W).
     \end{aligned}
\end{equation}
Here \(b\)  is the impact parameter, determined by \(b_{1}\), \(b_{2}\), and \(\phi\). The factor \(P_{\mathrm{NH}}(b)\) is probability that no hadronic interaction occurs at that impact parameter, thereby selecting genuinely ultra-peripheral events.

Due to the very low photon virtualities in heavy-ion UPC, as discussed in Sec.~\ref{sec::Intro}, the equivalent photons can be treated as quasi-real and their transverse momenta can be safely neglected. Consequently, the energies of the two photons can be determined by the invariant mass \(M\) and rapidity \(Y\) of the final state \(X\), such as a dilepton or di-hadron pair:
\begin{equation}
    \begin{cases}
     \omega_{1,2} = \frac{M}{2} e^{\pm Y}, \\
     Y = \frac{1}{2} \ln \frac{\omega_{1}}{\omega_{2}}.
    \end{cases}
     \label{eq:M_Y_omega}
\end{equation}
With this approximation, the photon-photon production differential cross-section can be expressed as:

\begin{equation}
    \begin{aligned}
        \frac{d\sigma_{A_{1} + A_{2} \rightarrow A_{1} + A_{2} + X}}{dM dY} &= \frac{M}{2} \int_{0}^{\infty} 2\pi b_{1} db_{1} \int_{0}^{\infty} 2\pi b_{2} db_{2} \\
        &\quad \times \int_{0}^{2\pi} \frac{d\phi}{2\pi} N(\omega_{1}, \mathbf{b}_{1}) N(\omega_{2}, \mathbf{b}_{2}) \\
        &\quad \times P_{\mathrm{NH}}(b) \sigma_{\gamma \gamma \rightarrow X}(W).
    \end{aligned}
    \label{eq:product_cross_section}
\end{equation}
By considering a specific range of rapidity as defined by the experiment, the cross section can be generated for different invariant mass ranges.

\subsection{Equivalent photon flux}

From EPA with modelled using the Weizsäcker–William method, the photon flux can be calculated from a classical photon flux distribution through considering the energy flux through an infinitesimal element of the transverse plane instead of the full transverse plane \cite{KRAUSS1997503}:
\begin{equation}
\begin{aligned}
N(\omega,r_{\perp}) 
&= \frac{Z^{2}\alpha_{QED}}{\pi^{2}\omega} \\
&\quad\times \left| \int_{0}^{\infty} dk_{\perp} \, k_{\perp}^{2} 
\frac{F\left(k_{\perp}^{2} + \left(\frac{\omega}{\gamma}\right)^{2} \right)}
{k_{\perp}^{2} + \left(\frac{\omega}{\gamma}\right)^{2}} 
J_{1}(r_{\perp}k_{\perp}) \right|^{2},
\label{eq:photon_flux}
\end{aligned}
\end{equation}
where \(\omega\) is the photon energy, \(r_{\perp}\) is the distance from the collided nucleus in beam view, \(k_{\perp}\) is the transverse part of photon frequency, \(\gamma\) is the Lorentz factor of the nucleus , \(Z\) is the proton number of the nucleus, \(\alpha_{QED}\) is the coupling constant of QED, \(J_{1}\) is the Bessel function of the first kind and \(F\) means electromagnetic form factor. Here \(\frac{\omega}{\gamma}\) is the longitudinal part of the photon frequency. In the laboratory frame, the photon energy has a cutoff of \( \omega_{\text{max}} \approx \gamma \hbar c/ R \), determined by the size of the nucleus. In the rest frame of the target nucleus, this cutoff is further boosted to approximately 500 GeV at RHIC and 1 PeV (1000 TeV) at the LHC \cite{UPCreview2}.

In our calculation, when the distance between the photon’s trajectory and the nuclear centre is smaller than the nuclear radius, we model the nuclear charge density \(\rho(r)\) with the Woods–Saxon (two-parameter Fermi) distribution \cite{woods1954}, which is widely used to describe the spatial distribution of nucleons inside a nucleus.  Its functional form reads  
\begin{equation}
  \rho(r)=\frac{\rho_{0}}{1+\exp\!\bigl[(r-R_{\mathrm{WS}})/d\bigr]},
  \label{eq:WoodSaxon}
\end{equation}
where \(\rho_{0}\) is a normalisation constant, \(R_{\mathrm{WS}}\) is the half-density radius, and \(d\) denotes the surface “skin” thickness.  
We adopt the empirical Woods–Saxon parameters extracted from elastic electron-scattering data~\cite{2pf_data}, with \(R_{\mathrm{WS}} = 6.38~\mathrm{fm}\) and \(d = 0.535~\mathrm{fm}\) for gold nuclei, and \(R_{\mathrm{WS}} = 6.62~\mathrm{fm}\) and \(d = 0.546~\mathrm{fm}\) for lead nuclei.

The corresponding nuclear electromagnetic form factor, defined as the Fourier transform of \(\rho(r)\), is  
\begin{equation}
  F(\mathbf{q}) = \int \rho(r)\,e^{i\mathbf{q}\cdot\mathbf{r}}\,
  \mathrm{d}^{3}\mathbf{r} ,
  \label{eq:FormFactor}
\end{equation}
with \(\mathbf{q}\) the momentum transfer. This is precisely the electromagnetic form factor $F$ entering Eq.~\eqref{eq:photon_flux}. 
Here, we evaluate it numerically by introducing a radial cut-off, beyond which $\rho(r)$ becomes negligibly small. A rigorous solution can be found in Eq. (17) of Ref.~\cite{Hua_sheng}.

For a "point-like" nuclear charge distribution (\(r_{\perp}> R_{\mathrm{WS}}\)), Eq.~\eqref{eq:photon_flux} simplifies to \cite{KRAUSS1997503}
\begin{equation}
  N(\omega,r_{\perp})\;=\;\frac{\mathrm{d}^{3}N}{\mathrm{d}\omega\,\mathrm{d}^{2}r_{\perp}}
  \;=\;
  \frac{Z^{2}\alpha_{QED}}{\pi^{2}\,\omega\,r_{\perp}^{2}}\;
  x^{2}\,K_{1}^{2}(x),
  \label{eq:photon_flux_simplified}
\end{equation}
with \(x=\omega r_{\perp}/\gamma \hbar c\) and \(K_{1}\) the modified Bessel function of the second kind.
As the photon flux quickly approaches the point-like limit for \(r_{\perp}>R_{\mathrm{WS}}\), one can therefore use the photon flux calculated with the Woods--Saxon charge distribution at small \(r_{\perp}\) and that with the point-like distribution at large \(r_{\perp}\), which greatly simplifies the computation without sacrificing numerical accuracy.

At RHIC, gold nuclei are accelerated to \(\sqrt{s_{NN}} = 200~\mathrm{GeV}\), corresponding to a Lorentz factor \(\gamma = 106.6\).  
At LHC, lead nuclei can reach \(\sqrt{s_{NN}} = 5.36~\mathrm{TeV}\), corresponding to a Lorentz factor \(\gamma = 2855.5\).  
This substantial difference in boost leads to important distinctions in the equivalent photon flux.  
In Eq.~\eqref{eq:photon_flux_simplified}, the flux falls off with the argument \( x = \omega r_\perp / \gamma \hbar c\), and hence a larger \( \gamma \) reduces \( x \), resulting in less suppression at large \( r_\perp \).  
As a consequence, the photon flux at the LHC is not only more intense but also extends farther in the transverse plane compared to RHIC.  

This leads to an enhanced spatial reach for two-photon interactions at the LHC and reflects the broader kinematic coverage of photon-induced processes in high-energy heavy-ion collisions.

\begin{figure*}[tbp]
    \centering
    \begin{subfigure}[b]{0.48\textwidth}
        \includegraphics[width=\linewidth]{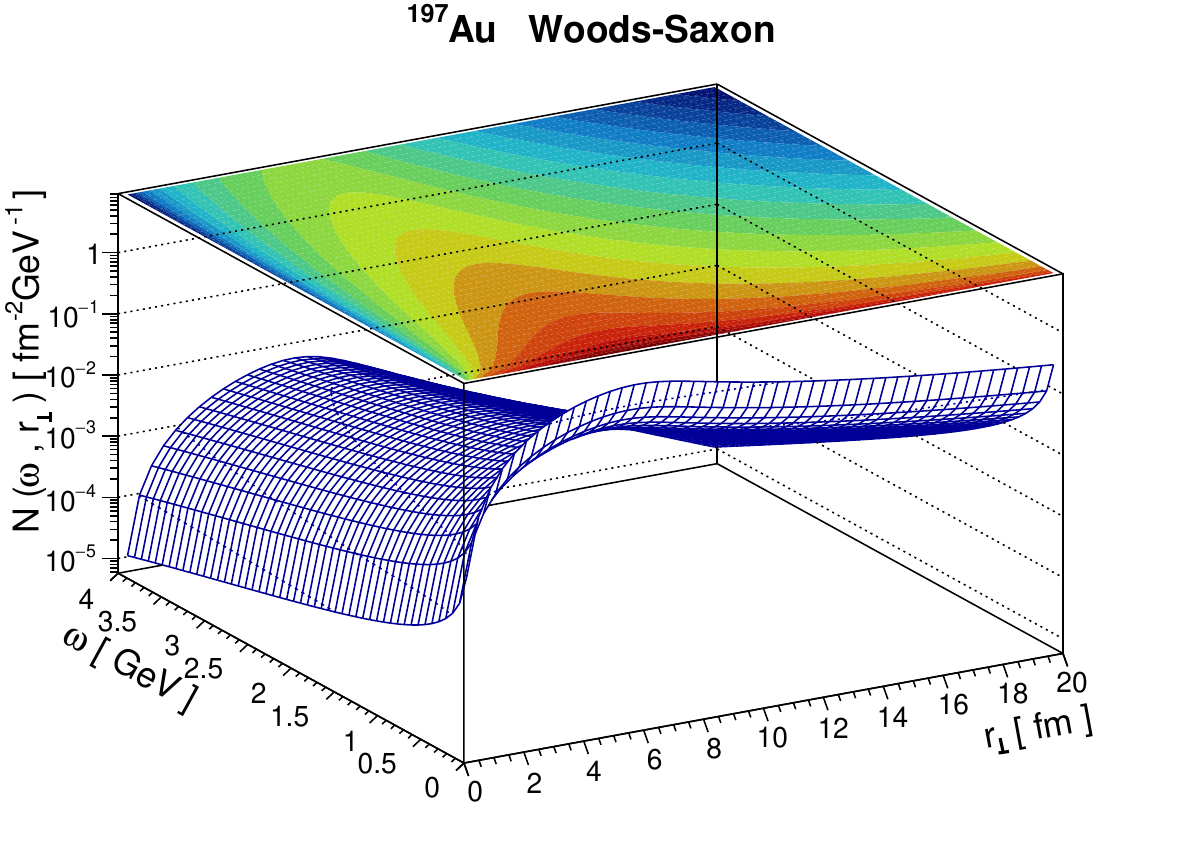}
        \caption{}
        \label{fig:au_photonflux}
    \end{subfigure}
    \hfill
    \begin{subfigure}[b]{0.48\textwidth}
        \includegraphics[width=\linewidth]{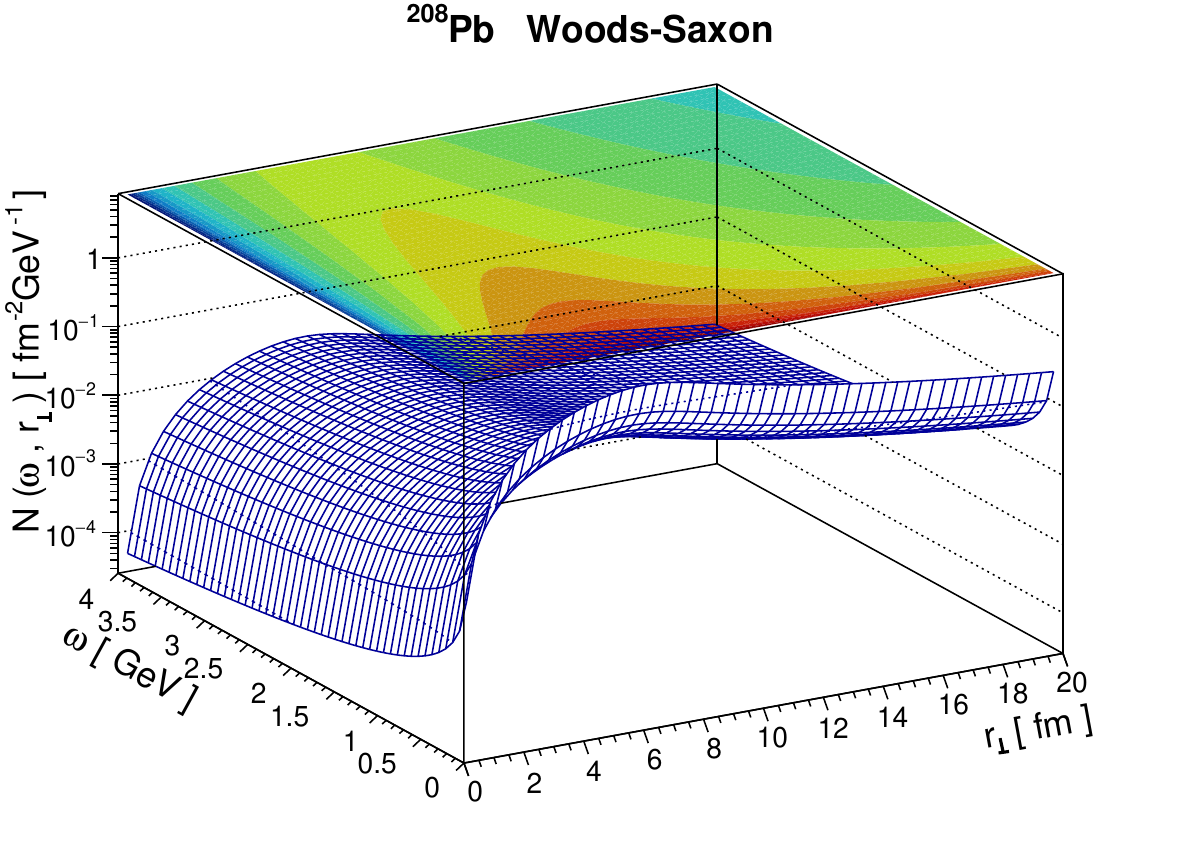}
        \caption{}
        \label{fig:pb_photonflux}
    \end{subfigure}
   \caption{Two-dimensional distributions of the photon flux as functions of transverse distance \(r_{\perp}\) from the nuclear center and photon energy \(\omega\), viewed along the beam direction. Calculations are performed using Woods–Saxon nuclear density profiles for (a) \(^{197}\mathrm{Au}\) nuclei in Au+Au collisions at \(\sqrt{s_{NN}} = 200\,\mathrm{GeV}\) at RHIC, and (b) \(^{208}\mathrm{Pb}\) nuclei in Pb+Pb collisions at \(\sqrt{s_{NN}} = 5.36\,\mathrm{TeV}\) at the LHC.}
    \label{fig:photon_flux_woodsaxon}
\end{figure*}

Using the Woods–Saxon parameters quoted above together with Eq.~\eqref{eq:photon_flux}, we evaluate the photon flux \(n(\omega, r_{\perp})\) as a function of the transverse distance \(r_{\perp}\) from the nuclear center (impact parameter) and the photon energy \(\omega\) in the nuclear rest frame. The resulting two-dimensional distributions are shown in Fig.~\ref{fig:au_photonflux} for RHIC and Fig.~\ref{fig:pb_photonflux} for LHC. As expected, the photon flux decreases with increasing $\omega$, following approximately a bremsstrahlung-like power-law behavior (roughly $\propto 1/\omega$).

\subsection{Probability of no hadronic interaction}
\label{subsec:P0H}

The Glauber model provides a convenient framework to estimate the probability that no hadronic occurs at a given impact parameter \(b\).  
Denoting this probability as \(P_{\mathrm{NH}}(b)\), one expects
\(P_{\mathrm{NH}}(b)\!\to\!0\) for central collisions \((b\lesssim2R)\) and  
\(P_{\mathrm{NH}}(b)\!\to\!1\) when the nuclei barely overlap \((b\gg2R)\).  
It is related to the complementary probability of hadronic interaction, \(P_{\mathrm{H}}(b)\), through
\begin{equation}
  P_{\mathrm{NH}}(b) \;=\; 1 - P_{\mathrm{H}}(b).
  \label{eq:Poh}
\end{equation}

For colliding nuclei with mass numbers \(A\) and \(B\), the standard Glauber form reads \cite{Miller2007}
\begin{align}
  P_{\mathrm{H}}(b) 
  &= 1-\Bigl[1-\sigma_{\mathrm{NN}}^{\mathrm{inel}}\,
             \hat{T}_{AB}(b)\Bigr]^{AB} \notag\\
  &\simeq 1-\exp\!\Bigl[-\sigma_{\mathrm{NN}}^{\mathrm{inel}}\,
            \hat{T}_{AB}(b)\Bigr],
  \label{eq:PH_Glauber}
\end{align}
where \(\sigma_{\mathrm{NN}}\) is the inelastic nucleon--nucleon cross-section and  
\(\hat{T}_{AB}(b)\) is the nuclear overlap function.
The thickness function of a single nucleus is defined as the line integral of its density profile:
\begin{equation}
  T_{A}(\mathbf{s}) \;=\; \int_{-\infty}^{\infty}\!\rho_{A}(\mathbf{s},z)\,
                           \mathrm{d}z,
  \label{eq:TA_def}
\end{equation}
where the density \(\rho_{A}\) is normalised to the mass number \(A\).
The overlap function is the convolution of the individual thickness functions,
\begin{align}
  \hat{T}_{AB}(b) 
  &= \int \mathrm{d}^{2}\mathbf{s}\,
     T_{A}(\mathbf{s})\,
     T_{B}(\mathbf{s}-\mathbf{b}) ,
\end{align}
with \(\mathbf{b}\) the transverse impact-parameter vector, and the integration extending over the entire two-dimensional transverse plane.

The inelastic nucleon–nucleon cross-section $\sigma_{\mathrm{NN}}^{\mathrm{inel}}$ used in Eq.~\eqref{eq:PH_Glauber} is obtained by isospin averaging \cite{mehndiratta2017}:
\begin{equation}
\begin{aligned}
\sigma_{\mathrm{NN}}^{\mathrm{inel}}
&=
\frac{
      N_P N_T \,\sigma_{nn}^{\mathrm{inel}}
      + Z_P Z_T \,\sigma_{pp}^{\mathrm{inel}}
      + (Z_P N_T + N_P Z_T)\,\sigma_{np}^{\mathrm{inel}}
     }
     {A_P A_T},
\end{aligned}
\label{eq:sigmaNN_avg}
\end{equation}
where $Z_{P,T}$ ($N_{P,T}$) denote the proton (neutron) numbers of the projectile
and target nuclei, respectively.
Following common practice, we set
$\sigma_{nn}^{\mathrm{inel}} = \sigma_{pp}^{\mathrm{inel}}$.
The numerical values of $\sigma_{pp}^{\mathrm{inel}}$ and
$\sigma_{np}^{\mathrm{inel}}$ are taken from PDG~2017~\cite{PDG2017}, with
$\sigma_{pp}^{\mathrm{inel}} = \sigma_{pp}^{\mathrm{total}} -
\sigma_{pp}^{\mathrm{el}}$ and
$\sigma_{np}^{\mathrm{inel}} = \sigma_{np}^{\mathrm{total}} -
\sigma_{pp}^{\mathrm{el}}$.
Here, $\sigma_{pp}^{\mathrm{el}}$ is used as a proxy for
$\sigma_{np}^{\mathrm{el}}$, since direct experimental data for
$\sigma_{np}^{\mathrm{el}}$ are not available.

With these inputs, we obtain
$\sigma_{\mathrm{NN}}^{\mathrm{inel}} = 40.83~\mathrm{mb}$ for Au+Au collisions at
$\sqrt{s_{\mathrm{NN}}}=200~\mathrm{GeV}$, corresponding to
$\sigma_{pp}^{\mathrm{inel}} = 40.69~\mathrm{mb}$ and
$\sigma_{np}^{\mathrm{inel}} = 40.99~\mathrm{mb}$.
For Pb+Pb collisions at $\sqrt{s_{\mathrm{NN}}}=5.36~\mathrm{TeV}$, we obtain
$\sigma_{\mathrm{NN}}^{\mathrm{inel}} = 66.66~\mathrm{mb}$, with
$\sigma_{pp}^{\mathrm{inel}} = 66.52~\mathrm{mb}$ and
$\sigma_{np}^{\mathrm{inel}} = 66.81~\mathrm{mb}$.
The near equality $\sigma_{pp}^{\mathrm{inel}} \simeq \sigma_{np}^{\mathrm{inel}}$ reflects the approximate isospin symmetry of the strong interactions at these energies.

\subsection{Mutual Coulomb Excitation in ultra-peripheral collisions}

Ultra–peripheral collisions of relativistic heavy ions provide exceptionally intense beams of quasi-real photons.  
For highly charged projectiles such as gold \((Z=79)\), the EPA predicts a photon flux large enough that, in addition to the exclusive two-photon process of interest (for example, \(p\bar p\) production), further photons can excite one or both nuclei, possibly accompanied by neutron emission.  
The dominated process in neutron emission is mutual Coulomb excitation (MCE): two or more photons populate the giant dipole resonance (GDR), and the subsequent de-excitation emits one or more neutrons \cite{Stelson1963}.  
GDR photo-nuclear cross-sections are known with high precision from dedicated measurements \cite{VEYSSIERE1970561,LEPRETRE1981237,CARLOS1984573,PhysRevD.5.1640,PhysRevD.7.1362,PhysRevLett.39.737,armstrong1972445}.

In collider experiments MCE serves as a convenient event tag.  
At RHIC–STAR, neutron signals in the Zero-Degree Calorimeters (ZDCs~\cite{RHIC_ZDC}) identify forward (\(+z\)) and backward (\(-z\)) nuclear break-up\cite{Klein2020,RHIC_neu_trigger}.

The average number of Coulomb excitations that emit at least one neutron, \(m_{Xn}(b)\), is obtained by folding the EPA flux with the photo-excitation cross-section \cite{BROZ2020107181}:
\begin{equation}
  m_{Xn}(b)
  = \int_{0}^{\infty} \mathrm{d}\omega \;
    N(\omega,b)\,
    \sigma_{\gamma A\to A^{*}}(\omega),
  \label{eq:mXn_def}
\end{equation}
where \(N(\omega,b)\) is given by Eq.\,\eqref{eq:photon_flux_simplified}.  
The calculation is confined to impact parameters \(b>12~\mathrm{fm}\); for smaller separations the probability of avoiding a hadronic interaction becomes negligible.

Because \(m_{Xn}(b)\) can exceed unity at high energies and small \(b\), a Poisson description is adopted \cite{BROZ2020107181,brandenburg2021}.  
The probability of absorbing exactly \(N\) photons is
\begin{equation}
  P^{(N)}(b)=
  \frac{[m_{Xn}(b)]^{N}\,
        e^{-m_{Xn}(b)}}{N!},
  \label{eq:Poisson_N}
\end{equation}
and the probability of at least one excitation is
\begin{equation}
  P_{Xn}(b)=1-e^{-m_{Xn}(b)}.
  \label{eq:PXn}
\end{equation}

Assuming the two nuclei excite independently, the joint probability for break-up modes \(i\) (projectile) and \(j\) (target) factorises:
\begin{equation}
  P_{ij}(b)=P_{i}(b)\,P_{j}(b).
  \label{eq:Pij}
\end{equation}

For \(\mathrm{Au}+\mathrm{Au}\) collisions at \(\sqrt{s_{NN}}=200~\mathrm{GeV}\) the STAR analysis focuses on neutron emission in the forward and backward directions.  
The shorthand \(X_{1}n\,X_{2}n\) denotes events with \(X_{1}\) neutrons from one nucleus and \(X_{2}\) neutrons from the other, while \(Xn\) refers to the emission of at least one neutron.

For Pb+Pb collisions at $\sqrt{s_{NN}}=5.36~\mathrm{TeV}$, the LHC experiments are able to distinguish different neutron-emission classes, while also providing cross sections that are inclusive with respect to neutron emission~\cite{CMS_2024_diel_NLO,CMS:2021dimuon,ATLAS_2020_dimuon,alice2013dielectron,atlas2023diele}. In this work, we focus on the neutron-inclusive results and perform our calculations accordingly.

\subsection{Elementary cross section of \(\gamma \gamma \rightarrow h\bar{h}\)}

Photon–photon production of hadron–antihadron pairs, \(\gamma\gamma\to h\bar h\), has been explored in \(e^{+}e^{-}\) colliders for several decades.  
Most measurements cover the central angular region \(|\cos\theta^{*}|<0.6\) in the \(h\bar h\) centre-of-mass frame.  
In ultra-peripheral heavy-ion collisions the same two-photon mechanism can create \(\pi^{+}\pi^{-}\), \(K^{+}K^{-}\), and \(p\bar p\) pairs, and the precision \(e^{+}e^{-}\) data provide indispensable input for modelling such production.

\begin{figure}[h]
  \centering
  \includegraphics[width=\linewidth]{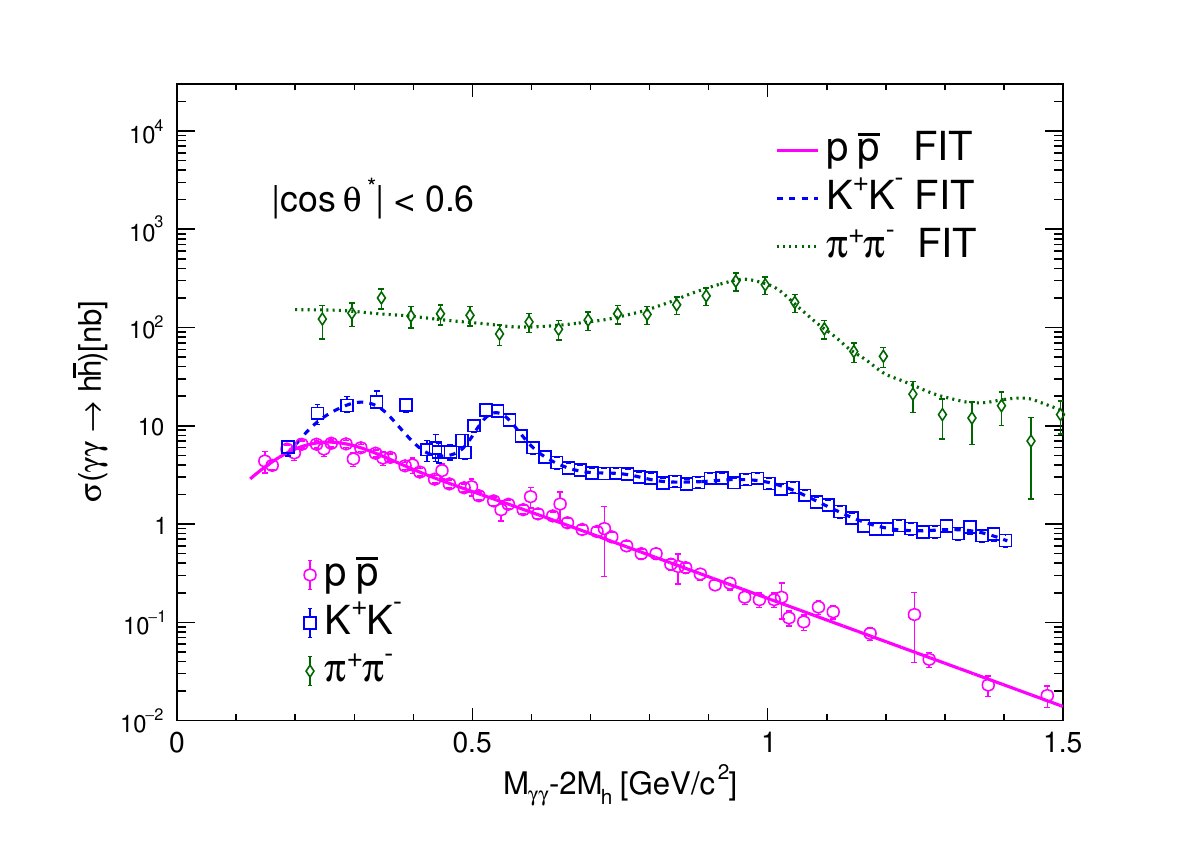}
  \caption{Cross-sections for \(\gamma\gamma\to h\bar h\) obtained in \(e^{+}e^{-}\) collisions:  
           \(\pi^{+}\pi^{-}\)~\cite{dipion_fit,dikaon_8_dipion_4},  
           \(K^{+}K^{-}\)~\cite{dikaon_3_argus,dikaon_4_belle2003}, and  
           \(p\bar p\)~\cite{Belle:2005fji,CLEO:1993ahu}.}
  \label{fig:gammagammadihadron}
\end{figure}

Comprehensive results exist for \(\gamma\gamma\to p\bar p\)~\cite{Belle:2005fji,CLEO:1993ahu,TASSO:1983mvj,JADE:1986ezn,OPAL:2002nhf,VENUS:1997can,L3:2003gyz,ARGUS:1988apz},  
\(\gamma\gamma\to K^{+}K^{-}\)~\cite{dikaon_1_LEP,dikaon_2_dipion_1_ALEPH,dikaon_3_argus,dikaon_4_belle2003,dikaon_5_dipion_2_belle2005,dikaon_6_dipion_3_CLEO,dikaon_7_TASSO,dikaon_8_dipion_4},  
and \(\gamma\gamma\to\pi^{+}\pi^{-}\)~\cite{dikaon_2_dipion_1_ALEPH,dikaon_5_dipion_2_belle2005,dikaon_6_dipion_3_CLEO,dikaon_8_dipion_4,dipion_5_SPEAR,dipion_6,dipion_7_MRAKII,dipion_8,dipion_9_Belle2007,dipion_11,dipion_12,dipion_13}.  
These datasets are displayed together with phenomenological fits in Fig.\,\ref{fig:gammagammadihadron}.  
For the ARGUS \(K^{+}K^{-}\) points~\cite{dikaon_3_argus}, the quoted values are extrapolated to \(|\cos\theta^{*}|<0.6\) using the angular distribution measured by Belle in the closest mass interval.

The ultra-peripheral heavy-ion cross-sections are evaluated by folding these \(\gamma\gamma\to h\bar h\) data with the equivalent-photon spectra of the colliding ions. 
In the high invariant-mass region, the existing $e^{+}e^{-}$ measurements are sparse and subject to large uncertainties, particularly for $M_{p\bar{p}} > 3.0~\mathrm{GeV}/c^{2}$. Therefore, in this study we conservatively limit our calculations to $M_{p\bar{p}} < 3.0~\mathrm{GeV}/c^{2}$, where the available experimental data provide more reliable constraints.

\subsection{Kinematics of the \(h\bar{h}\) pair}\label{Sec:PtspectrumMC}

The transverse-momentum distribution of the produced di-hadron pair is governed by the photon transverse momenta.  
The differential spectrum of a single photon is obtained by integrating the EPA flux over transverse coordinates \cite{brandenburg2021}:
\begin{equation}
  \frac{\mathrm{d}N}{\mathrm{d}k_{\perp}}
  =\frac{1}{2\pi^{2}}\,
    \frac{F^{2}\!\bigl(k_{\perp}^{2}+\omega^{2}/\gamma^{2}\bigr)\,
          k_{\perp}^{3}}
         {\bigl(k_{\perp}^{2}+\omega^{2}/\gamma^{2}\bigr)^{2}},
  \label{eq:Pt_distribution}
\end{equation}
where \(F\) is the nuclear electromagnetic form factor evaluated with the Woods–Saxon density.  
This approximation provides a sufficiently accurate description for the present study.

The pair transverse momentum is constructed as the vector sum of the two photon momenta,
\( \mathbf{P}_{\perp}= \mathbf{k}_{1\perp}+\mathbf{k}_{2\perp}\);  
the azimuthal directions of \(\mathbf{k}_{1\perp}\) and \(\mathbf{k}_{2\perp}\) are taken to be random. Although \(P_{\perp}\) is correlated with the pair invariant mass and rapidity through the photon energies, all spectra generated within this framework are mutually consistent for each specific type of particle pair within the STAR acceptance. Owing to the broader rapidity range covered by typical LHC detectors, the spectra depend on the specific rapidity interval analyzed.

The angular distribution \(\gamma\gamma\to h\bar h\) entering the calculation is taken directly from experimental data.  
The STAR analysis applies the cuts  
\(0.05<|Y^{h\bar h}|<0.5\),  
\(P^{h\bar h}_{T}<0.1\;\mathrm{GeV}/c\),  
\(P^{h,\bar h}_{T}>0.2\;\mathrm{GeV}/c\),  
and \(|\eta^{h,\bar h}|<0.9\).  
These criteria extend the centre-of-mass angular coverage beyond the usual \(|\cos\theta^{*}|<0.6\) window explored in most \(e^{+}e^{-}\) studies.  
However, the contribution from \(|\cos\theta^{*}|>0.6\) remains negligible, becoming noticeable only for pair rapidities below 0.2, where it is estimated not to exceed one percent.
For typical acceptance of LHC, the cuts applied are \(0 < |Y^{h\bar{h}}| < 2.0\),  
\(P^{h\bar{h}}_T < 0.1\,\mathrm{GeV}/c\),  
\(P^{h,\bar{h}}_T > 0.2\,\mathrm{GeV}/c\),  
and \(|\eta^{h,\bar{h}}| < 2.4\).  
Within the invariant mass range considered in our study, full coverage of the region \(|\cos\theta^{*}| < 0.6\) is challenging to achieve.  
For the portion with \(|\cos\theta^{*}| > 0.6\), we perform an extrapolation based on the known angular distributions.

\section{Results}
\label{sec:results}

\begin{figure*}[tbp]
    \centering
    \begin{subfigure}[b]{0.48\textwidth}
        \includegraphics[width=\linewidth]{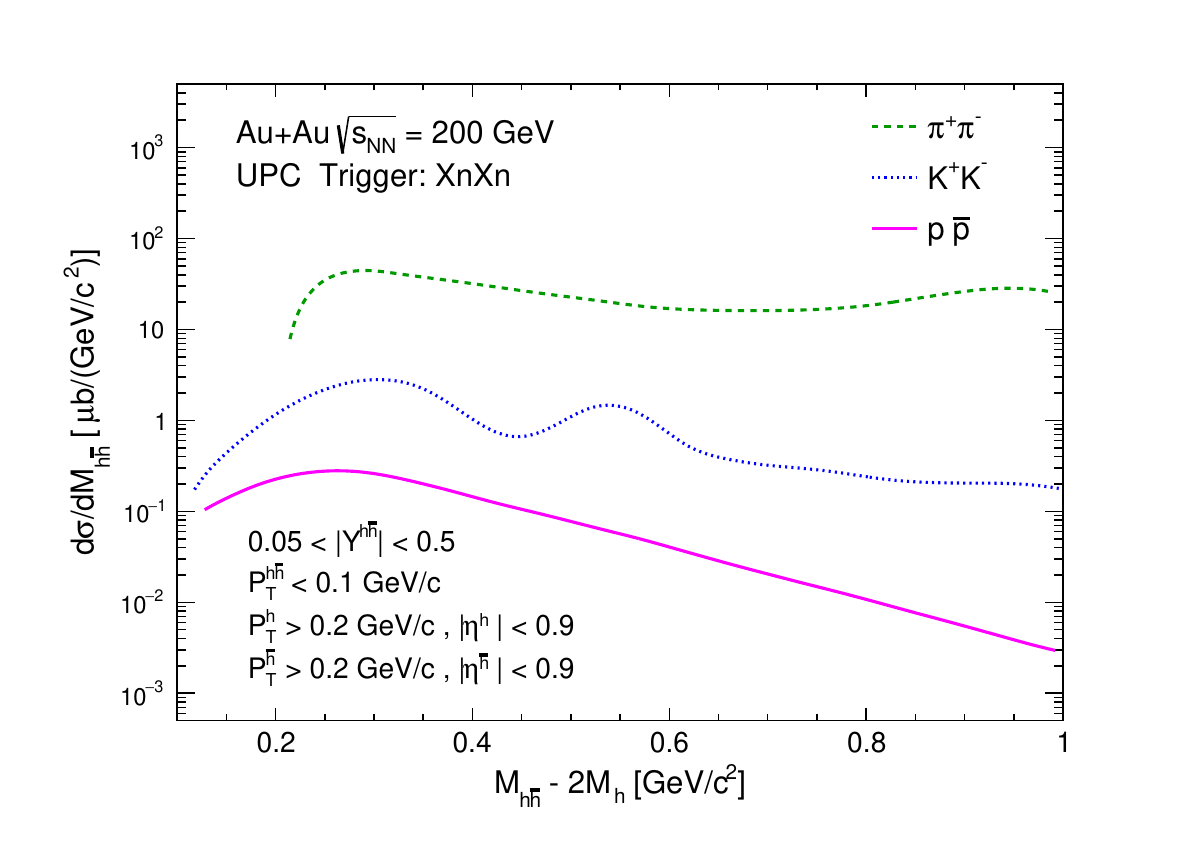}
        \caption{}
        \label{fig:dihardon_STAR}
    \end{subfigure}
    \hfill
    \begin{subfigure}[b]{0.48\textwidth}
        \includegraphics[width=\linewidth]{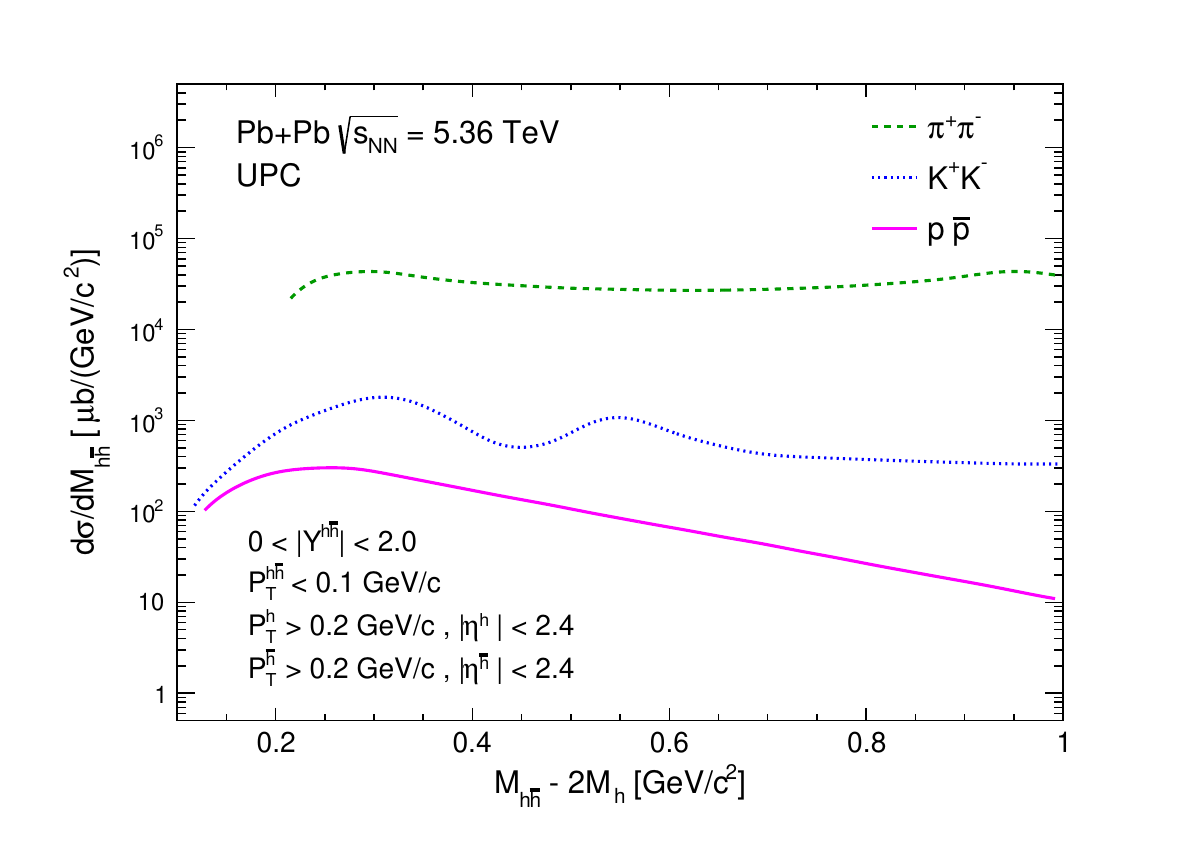}
        \caption{}
        \label{fig:dihardon_CMS}
    \end{subfigure}
   \caption{The production cross sections of dihadron pairs via the two-photon process in \(\mathrm{Au{-}Au}\) UPC at \(\sqrt{s_{NN}} = 200~\mathrm{GeV}\) within the STAR acceptance (a) and in \(\mathrm{Pb{-}Pb}\) UPC at \(\sqrt{s_{NN}} = 5.36~\mathrm{TeV}\) within typical acceptance of LHC (b), which are calculated based on fits to experimental data from electron--positron collisions for the \( K^+K^- \)\cite{dikaon_3_argus,dikaon_4_belle2003}, \( \pi^+\pi^- \)\cite{dipion_fit,dipion_7_MRAKII}, and \( p\bar{p} \)\cite{Belle:2005fji}.}
    \label{fig:dihardon}
\end{figure*}

\begin{figure*}[tbp]
    \centering
    \begin{subfigure}[b]{0.48\textwidth}
        \includegraphics[width=\linewidth]{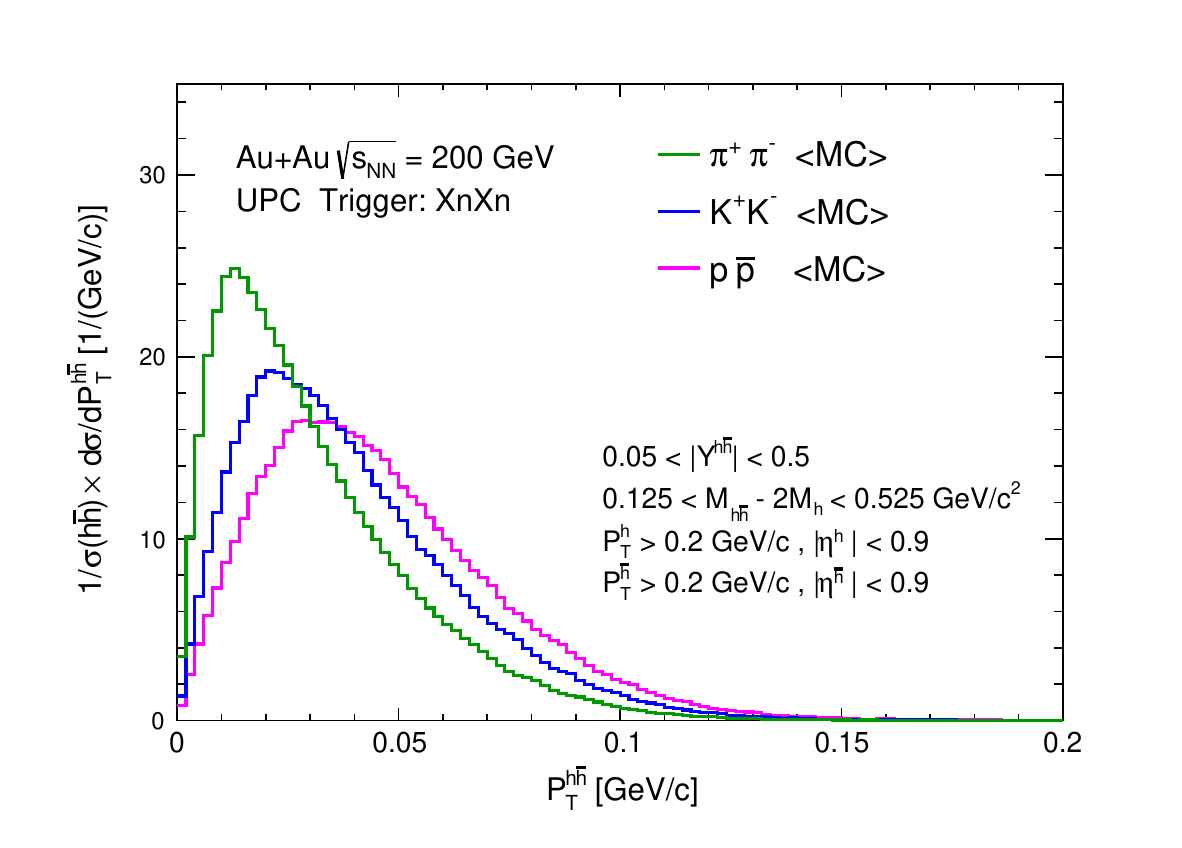}
        \caption{}
        \label{fig:Ptspectrum_STAR}
    \end{subfigure}
    \hfill
    \begin{subfigure}[b]{0.48\textwidth}
        \includegraphics[width=\linewidth]{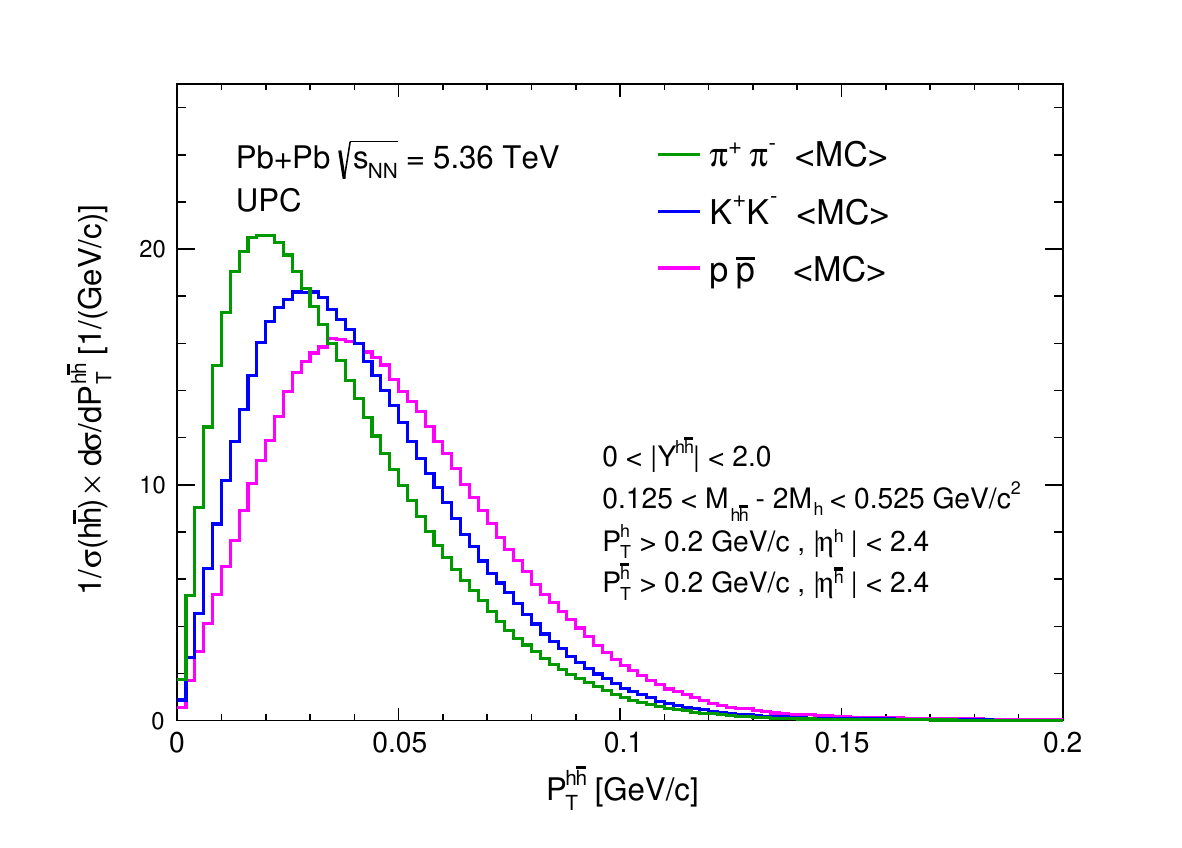}
        \caption{}
        \label{fig:Ptspectrum_CMS}
    \end{subfigure}
   \caption{The normalized transverse momentum (\(P^{h\bar{h}}_{T}\)) spectra, i.e. $1/\sigma(h\bar{h}) \times d\sigma/dP^{h\bar{h}}_{T}$, of \(\pi^{+}\pi^{-}\), \(K^{+}K^{-}\), and \(p\bar{p}\) pairs produced via the two-photon process in ultra-peripheral \(\mathrm{Au{-}Au}\) collisions at \(\sqrt{s_{NN}} = 200~\mathrm{GeV}\) within the STAR acceptance (a), and in \(\mathrm{Pb{-}Pb}\) collisions at \(\sqrt{s_{NN}} = 5.36~\mathrm{TeV}\) within typical acceptance of LHC (b), calculated using Monte Carlo simulations.}
    \label{fig:Ptspectrum}
\end{figure*}

Based on Eq.~\eqref{eq:product_cross_section}, the differential cross section for \(h\bar{h}\) production in ultra--peripheral \(\mathrm{Au{+}Au}\) collisions is obtained by folding the photon--photon fusion cross section with the equivalent photon flux of each nucleus.  The fusion cross section, \(\sigma_{\gamma\gamma\to h\bar{h}}\), is taken from experimental \(e^{+}e^{-}\) data at the corresponding invariant mass. For the photon flux we employ Eq.~\eqref{eq:photon_flux}, which uses a Woods--Saxon charge distribution inside the nuclear radius, and Eq.~\eqref{eq:photon_flux_simplified}, which assumes a point--like source outside.  The two parameterisations converge smoothly at large impact parameters.

To account for event selection, we quantify (i)~the probability of no hadronic interaction, \(P_{\mathrm{NH}}(b)\), via Eq.~\eqref{eq:Poh}, and (ii)~the probability for mutual Coulomb excitation accompanied by neutron emission (\(XnXn\) trigger) via Eqs.~\eqref{eq:mXn_def}--\eqref{eq:PXn}.  Both quantities depend on the impact parameter~\(b\).  With a Woods--Saxon nuclear density and the average nucleon--nucleon cross section, \(P_{\mathrm{NH}}(b)\to 0\) for \(b\lesssim 12~\mathrm{fm}\) and \(P_{\mathrm{NH}}(b)\to 1\) for \(b\gtrsim 20~\mathrm{fm}\).   The mean number of Coulomb excitations leading to Giant Dipole Resonance neutron emission is computed from the photon flux and the photo--excitation cross section, assuming Poisson statistics.  For the no-hadronic interaction region corresponding to \(b > 12~\mathrm{fm}\), we treat each nucleus as point--like in the excitation integral, consistent with their physical separation.

The overall detection acceptance under the STAR acceptance is evaluated with a dedicated Monte--Carlo (MC) simulation.  The transverse--momentum spectrum of the produced \(p\bar{p}\) pair is governed by the transverse momenta of the two quasi--real photons, distributed according to Eq.~\eqref{eq:Pt_distribution} and assumed to be uncorrelated in the transverse plane.  The polar--angle distribution of the subprocess \(\gamma\gamma\to h\bar{h}\) is sampled from experimental data in the region \(|\cos\theta^{*}|<0.6\).
For typical acceptance of LHC, we employ the same MC simulation framework; however, the angular coverage is extended by extrapolating the distribution from \(|\cos\theta^{*}| < 0.6\) to the full range \(|\cos\theta^{*}| < 1.0\). For the $\pi^{+}\pi^{-}$~\cite{dipion_7_MRAKII} and $K^{+}K^{-}$~\cite{dikaon_4_belle2003} channels, the angular distributions are extrapolated using partial-wave fits, while for the $p\bar{p}$~\cite{Belle:2005fji} channel a simple third-order polynomial parameterization is used as the description of the angular distribution in the extrapolated region.

Fig.~\ref{fig:dihardon} displays the calculated production cross sections for \(\pi^{+}\pi^{-}\), \(K^{+}K^{-}\) and \(p\bar{p}\) pairs in \(\mathrm{Au{+}Au}\) ultra--peripheral collisions at \(\sqrt{s_{NN}}=200~\mathrm{GeV}\) and \(\mathrm{Pb{+}Pb}\) ultra--peripheral collisions at \(\sqrt{s_{NN}}=5.36~\mathrm{TeV}\). The kinematic requirements for the \(\mathrm{Au{+}Au}\) system within the STAR acceptance are  
\(0.05 < |Y^{h\bar{h}}| < 0.5\),  
\(P^{h\bar{h}}_T < 0.1\,\mathrm{GeV}/c\),  
\(P^{h,\bar{h}}_T > 0.2\,\mathrm{GeV}/c\),  
and \(|\eta^{h,\bar{h}}| < 0.9\),  
and for the \(\mathrm{Pb{+}Pb}\) system within typical acceptance of LHC, the cuts are  
\(0 < |Y^{h\bar{h}}| < 2.0\),  
\(P^{h\bar{h}}_T < 0.1\,\mathrm{GeV}/c\),  
\(P^{h,\bar{h}}_T > 0.2\,\mathrm{GeV}/c\),  
and \(|\eta^{h,\bar{h}}| < 2.4\),  
all of which are consistent with those detailed in Sec.~\ref{Sec:PtspectrumMC}.

The resulting cross sections are of order \(\mathcal{O}(\mu\mathrm{b})\) for STAR.  Within the corresponding cuts the STAR detector acceptance is \(\sim60\,\%\) for \(p\bar{p}\) pairs and \(\sim85\,\%\) for \(\pi^{+}\pi{-}\) and \(K^{+}K^{-}\) pairs near the peak within \(|\cos\theta^{*}|<0.6\), providing substantial kinematic
coverage. A clear mass--threshold hierarchy is visible: \(\sigma_{\pi^{+}\pi^{-}} \gg \sigma_{K^{+}K^{-}} \gg \sigma_{p\bar{p}}\), with each successive channel suppressed by roughly one order of magnitude.

The resulting cross sections are of order \(\mathcal{O}(\mathrm{mb})\) for LHC experiments, representing an increase of roughly three orders of magnitude compared to the STAR results. This enhancement primarily arises from the combined effects of the higher center--of--mass energy and the removal of the \(XnXn\) trigger requirement. Within the corresponding cuts, the LHC detector acceptance is \(\sim100\,\%\) for both \(p\bar{p}\) pairs and \(\pi^{+}\pi^{-}\), \(K^{+}K^{-}\) pairs near the peak within \(|\cos\theta^{*}|<0.6\).  A notable point is that, unlike the STAR acceptance---which can be almost fully covered by the \(|\cos\theta^{*}|<0.6\) region relevant for \(e^{+}e^{-}\) collisions---the LHC calculation receives a significant contribution from the \(|\cos\theta^{*}|>0.6\) region.

The systematic uncertainties of our calculations are estimated to arise from three main sources:

\noindent
(i) \textit{Uncertainties in the nuclear parameters.}
To assess this effect, we repeat the calculations using an alternative set of nuclear parameters~\cite{Shou_nuclear_paranew}, with \(R_{\mathrm{WS}} = 6.42~\mathrm{fm}\) and \(d = 0.41~\mathrm{fm}\) for gold nuclei, and \(R_{\mathrm{WS}} = 6.66~\mathrm{fm}\) and \(d = 0.45~\mathrm{fm}\) for lead nuclei. The resulting deviations are found to be below $1\%$ for both STAR and LHC kinematics.

\noindent
(ii) \textit{Uncertainties of the $\gamma\gamma \to h\bar{h}$ cross sections measured in $e^{+}e^{-}$ collisions.}
We assign a systematic uncertainty of $2\%$ for the $K^{+}K^{-}$ and $p\bar{p}$ channels, and $1\%$ for the $\pi^{+}\pi^{-}$ channel, based on the experimental uncertainties of the measured $\gamma\gamma \to h\bar{h}$ cross sections in $e^{+}e^{-}$ collisions (the fit for $\pi^{+}\pi^{-}$ channel is taken from Ref.~\cite{dipion_fit}).

\noindent
(iii) \textit{Uncertainties associated with the angular distributions.}
In $e^{+}e^{-}$ experiments, the angular distributions are typically provided bin by bin in the pair invariant mass, and differences between neighboring mass bins can lead to non-smooth behavior in the calculated cross sections. To quantify this effect, we compare the results obtained using the original, non-smooth angular distributions with those obtained after applying the smoothing procedure adopted in this work. This comparison effectively accounts for the variations of the angular distributions between adjacent invariant-mass bins and is taken as an estimate of the corresponding systematic uncertainty, which amounts to less than $3\%$ for the STAR kinematics and up to $10\%$ for the LHC kinematics. The larger angular-distribution uncertainty at the LHC mainly originates from the significantly wider acceptance, which includes a non-negligible fraction of the $|\cos\theta^{*}| > 0.6$ region and is therefore strongly correlated with the extrapolation procedure employed in our analysis.

For the STAR kinematics, the total systematic uncertainty is found to be below $4\%$, with the dominant contribution arising from the angular distributions. For the LHC kinematics, the total systematic uncertainty is below $10\%$, again dominated by the uncertainty associated with the angular distributions.

To facilitate comparison with the recent STAR measurement, we also compute the \(h\bar{h}\) transverse--momentum spectrum under the tighter mass window \(0.125<M_{h\bar{h}}-2M_{h}<0.525~\mathrm{GeV}/c^{2}\).  Fig.~\ref{fig:Ptspectrum_STAR} shows the normalized \(P^{h\bar{h}}_{T}\) spectrum , i.e. $1/\sigma(h\bar{h}) \times d\sigma/dP^{h\bar{h}}_{T}$, obtained from the MC within the STAR acceptance.
Nearly all yield resides below \(P^{h\bar{h}}_{T}<0.15~\mathrm{GeV}/c\), as expected for coherent two--photon production.
For LHC, we apply the same mass window \(0.125 < M_{h\bar{h}} - 2M_h < 0.525~\mathrm{GeV}/c^2\). Fig.~\ref{fig:Ptspectrum_CMS} shows the corresponding normalized \(P^{h\bar{h}}_T\) spectrum within the LHC acceptance.
The normalized $P^{h\bar{h}}_{T}$ spectrum is strongly correlated with the invariant mass and rapidity ranges of the hadron pair, as dictated by Eqs.~\eqref{eq:M_Y_omega} and~\eqref{eq:Pt_distribution}.
Due to the wider pair rapidity coverage in LHC, the  normalized \(P^{h\bar{h}}_T\) spectrum is slightly broadened compared to that in STAR, though nearly all the yield still resides below \(P^{h\bar{h}}_T < 0.15~\mathrm{GeV}/c\), and it is essentially insensitive to the uncertainties from the extrapolation of the angular distribution. 
For the absolute $P^{h\bar{h}}_{T}$ spectrum, $d\sigma/dP^{h\bar{h}}_{T}$, the systematic uncertainty is dominated by the uncertainty of the total cross section, while the spectral shape, defined by $1/\sigma(h\bar{h}) \times d\sigma/dP^{h\bar{h}}_{T}$, remains stable. The systematic uncertainties of the total cross section follow the same considerations as discussed above.

\begin{figure}[h]
    \centering
    \includegraphics[width=1.0\linewidth]{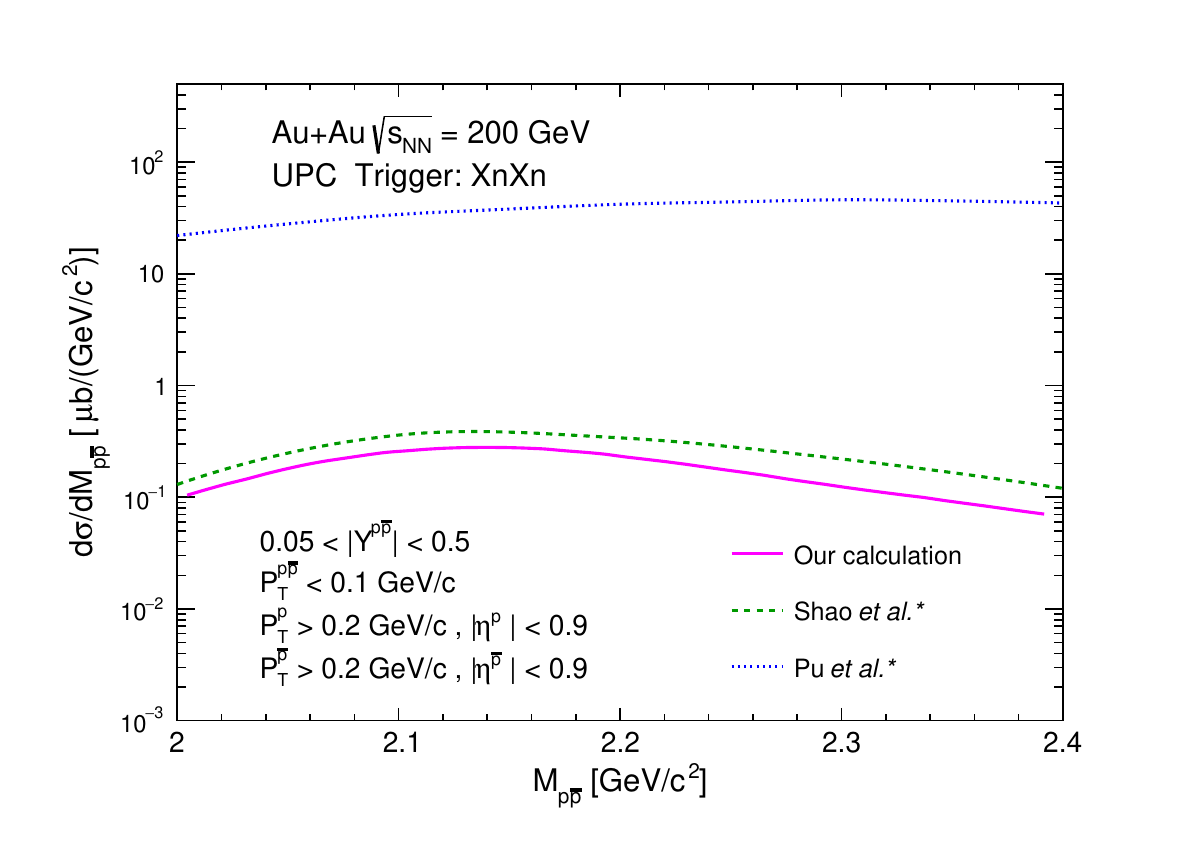} 
    \caption{Calculation of the \( p\bar{p} \) production cross section via two-photon process in Au+Au UPC collisions at \( \sqrt{s_{NN}} = 200~\mathrm{GeV} \), compared with theoretical predictions from Pu et al.~\cite{ppbar_Pu} and Shao et al.~\cite{ppbar_Shao}.}
    \label{fig:compare_others}
\end{figure}

In addition, we have compared our calculation of the \( p\bar{p} \) production cross section via photon–photon fusion in \(\mathrm{Au{+}Au}\) UPC at RHIC, under the kinematic conditions of the recent STAR report, with other theoretical studies of the same process. Our result shows excellent agreement with the prediction by Shao et al.~\cite{ppbar_Shao}, while the cross section obtained by Pu et al.~\cite{ppbar_Pu} is two to three orders of magnitude larger, as shown in Fig.~\ref{fig:compare_others}.

Previous UPC measurements of lepton--pair production, modelled within the same
EPA formalism, reproduce the data to high accuracy when described by
leading--order QED calculations at RHIC~\cite{Star2004:e+e-,Star2017:e+e-},
and by next--to--leading--order QED calculations at the LHC~\cite{alice2013dielectron,atlas2023diele,CMS:2021dimuon,Shao_NLO,CMS_2024_diel_NLO,ATLAS_2020_dimuon}.
In contrast, $h\bar{h}$ production requires $\gamma\gamma \to h\bar{h}$ cross sections extracted from $e^{+}e^{-}$ experiments. As discussed in Sec.~\ref{sec::Intro}, the photon virtualities differ between the two environments, being typically $Q^{2}\sim10^{-2}~\mathrm{GeV}^{2}$ in $e^{+}e^{-}$ collisions, while they are limited to $Q^{2}\lesssim10^{-3}~\mathrm{GeV}^{2}$ in heavy--ion UPC for gold and lead nuclei. The prevailing view is that such differences can be neglected; however, it has been argued that the resulting photon virtuality effects may still have a non-negligible impact on certain observables~\cite{Qdependence}. Aside from the difference in photon virtuality, we consider the two-photon environments in \( e^{+}e^{-} \) and heavy-ion UPC to be physically equivalent. From a theoretical standpoint, the underlying production mechanism should be identical in both cases. Therefore, provided that the photon virtuality does not lead to significant modifications of the cross section, our calculation (based on the fitted results from \( e^{+}e^{-} \) data) should correspond directly to the differential cross section for the two-photon process in UPC.
The present calculation therefore provides a unique test of photon-virtuality corrections and offers new constraints on electromagnetic interactions and baryon structure in the nearly real-photon regime.

\section{Summary}

Using the Equivalent Photon Approximation, we calculated baseline cross sections for exclusive \(\pi^{+}\pi^{-}\), \(K^{+}K^{-}\), and \(p\bar p\) production in \(\sqrt{s_{NN}}=200\;\text{GeV}\) Au+Au and \(\sqrt{s_{NN}}=5.36\;\text{TeV}\) Pb+Pb ultra-peripheral collisions, incorporating photon fluxes based on EPA, impact-parameter–dependent hadronic-suppression and Coulomb-excitation probabilities, and STAR/LHC kinematic cuts. The predicted yields are at the \(\mu\text{b}\) level for Au+Au and at the mb level for Pb+Pb, with a clear hierarchy \(\sigma_{\pi^{+}\pi^{-}}\gg\sigma_{K^{+}K^{-}}\gg\sigma_{p\bar p}\). In both systems, the transverse-momentum spectrum of the di-hadron pair peaks sharply below \(0.15~\mathrm{GeV}/c\). Within the kinematic region relevant to our study, the transverse-momentum spectra exhibit a consistent ordering in their peak positions: the \( \pi^+\pi^- \) channel peaks at the lowest transverse momentum, followed by \( K^+K^- \), and finally \( p\bar{p} \). These results provide a unified reference for di-hadron two-photon production in heavy-ion collisions, offering immediate benchmarks for upcoming STAR and LHC measurements.

\section*{Acknowledgement}
This work is supported in part by the National Key Research and Development Program of China under Contract No. 2022YFA1604900 the National Natural Science Foundation of China (NSFC) under Contract No. 12422510. W. Zha is supported by Anhui Provincial Natural Science Foundation No. 2508085JX002, Youth Innovation Promotion Association of Chinese Academy of Sciences and the Chinese Academy of Sciences (CAS) under Grants No. YSBR-088. X. Li is supported by National Natural Science Foundation of China (NSFC) under grant No. 124B2103.

\bibliographystyle{unsrt}
\bibliography{reference}

\end{document}